\documentclass[twocolumn]{aastex631}

\usepackage{url}

\usepackage{amsmath}
\usepackage{float}
\usepackage{rotating}
\usepackage{hyperref}
\graphicspath{{./}{Figures/}}

\shorttitle{SOLES I: The Spin-Orbit Alignment of K2-140 b}
\shortauthors{Rice et al.}

\begin{document}
\title{Stellar Obliquities in Long-period Exoplanet Systems (SOLES) I: The Spin-Orbit Alignment of K2-140 b}

\author[0000-0002-7670-670X]{Malena Rice}
\altaffiliation{NSF Graduate Research Fellow}
\affiliation{Department of Astronomy, Yale University, New Haven, CT 06511, USA}

\author[0000-0002-7846-6981]{Songhu Wang}
\affiliation{Department of Astronomy, Indiana University, Bloomington, IN 47405, USA}

\author[0000-0001-8638-0320]{Andrew W. Howard}
\affiliation{Department of Astronomy, California Institute of Technology, Pasadena, CA 91125, USA}

\author[0000-0002-0531-1073]{Howard Isaacson}
\affiliation{Department of Astronomy, University of California Berkeley, Berkeley CA 94720, USA}
\affiliation{Centre for Astrophysics, University of Southern Queensland, Toowoomba, QLD, Australia}

\author[0000-0002-8958-0683]{Fei Dai}
\affiliation{Division of Geological and Planetary Sciences, California Institute of Technology, Pasadena, CA 91125, USA}

\author[0000-0002-0376-6365]{Xian-Yu Wang}
\affiliation{National Astronomical Observatories, Chinese Academy of Sciences, Beijing 100012, China}
\affiliation{University of the Chinese Academy of Sciences, Beijing, 100049, China}

\author[0000-0001-7708-2364]{Corey Beard}
\affiliation{Department of Physics and Astronomy, University of California Irvine, Irvine, CA 92697, USA}

\author[0000-0003-0012-9093]{Aida Behmard}
\altaffiliation{NSF Graduate Research Fellow}
\affiliation{Division of Geological and Planetary Sciences, California Institute of Technology, Pasadena, CA 91125, USA}

\author[0000-0001-7676-6182]{Casey Brinkman}
\altaffiliation{NSF Graduate Research Fellow}
\affiliation{Institute for Astronomy, University of Hawai'i at Manoa, Honolulu, HI 96822, USA}

\author[0000-0003-3856-3143]{Ryan A. Rubenzahl}
\altaffiliation{NSF Graduate Research Fellow}
\affiliation{Department of Astronomy, California Institute of Technology, Pasadena, CA 91125, USA}

\author[0000-0002-3253-2621]{Gregory Laughlin}
\affiliation{Department of Astronomy, Yale University, New Haven, CT 06511, USA}

\correspondingauthor{Malena Rice}
\email{malena.rice@yale.edu}

\begin{abstract}
Obliquity measurements for stars hosting relatively long-period giant planets with weak star-planet tidal interactions may play a key role in distinguishing between formation theories for shorter-period hot Jupiters. Few such obliquity measurements have been made to date due to the relatively small sample of known wide-orbiting, transiting Jovian-mass planets and the challenging nature of these targets, which tend to have long transit durations and orbit faint stars. We report a measurement of the Rossiter-McLaughlin effect across the transit of K2-140 b, a Jupiter-mass planet with period $P=6.57$ days orbiting a $V=12.6$ star. We find that K2-140 is an aligned system with projected spin-orbit angle $\lambda=0.5\degr\pm9.7\degr$, suggesting a dynamically cool formation history. This observation builds towards a population of tidally detached giant planet spin-orbit angles that will enable a direct comparison with the distribution of close-orbiting hot Jupiter orbital configurations, elucidating the prevalent formation mechanisms of each group.
\end{abstract}


\section{Introduction} 
\label{section:intro}

The obliquity of a star, or the degree of alignment between the star's spin axis and its companion planets' net orbital angular momentum axis, provides crucial insights into the dynamical evolution of the surrounding system. Trends in stellar obliquity can constrain the prevalence of various processes crafting the observed distribution of extrasolar planets \citep[e.g.][]{winn2010hot, albrecht2012obliquities, winn2015}.

While the obliquities of many hot-Jupiter-hosting stars have been determined through measurements of the Rossiter-McLaughlin effect \citep{rossiter1924detection, mclaughlin1924some}, only a handful of these measurements have been made for systems with planets orbiting at larger distances from their host star, where star-planet tidal interactions are too weak to significantly influence the planets' orbital alignment. We refer to these wide-orbiting planets as ``tidally detached." Because the probability $p$ of transit for a given planet with semimajor axis $a$ falls as $p\propto 1/a$, fewer tidally detached transiting giant planets have been discovered to date as compared with closer-in hot Jupiters. Due to the volume-limited sample of bright stars, those tidally detached Jupiters that have been discovered are often challenging Rossiter-McLaughlin targets, orbiting stars too faint to obtain a sufficient number of high-resolution spectroscopic observations across a single transit with any but the largest existing ground-based telescopes. 

The vast majority of spin-orbit angle measurements have, consequently, been made for planets on tight orbits with $a/R_*\lesssim12$. However, the obliquities of stars hosting wider-separation giant planets may provide the evidence necessary to distinguish between the various proposed formation mechanisms for hot Jupiters \citep{winn2015, dawson2018origins}. Furthermore, in the classical quiescent formation framework for hot and warm Jupiters, the long tidal realignment timescales of these systems enable a direct measurement of the primordial dispersion in protoplanetary disk misalignments.

We present a Rossiter-McLaughlin measurement across the transit of K2-140 b with the High Resolution Echelle Spectrometer \citep[HIRES;][]{vogt1994hires} on the 10-meter Keck I telescope. K2-140 (EPIC 228735255) is a $V=12.6$ G5 star hosting a 1.019±0.070$\mathrm{M_J}$ planet, K2-140 b, with a $P=6.57-$day orbital period \citep{giles2018k2}. With $a/R_*=12.88$, K2-140 b is one of the widest-separation tidally detached planets to date with a measured spin-orbit angle. This is the first measurement in our Stellar Obliquities in Long-period Exoplanet Systems (SOLES) survey designed to expand the sample of Rossiter-McLaughlin measurements for wide-separation exoplanets.

K2-140 b was first characterized using photometry from the \textit{K2} mission \citep{howell2014k2}, a re-purposed extension of the \textit{Kepler} mission \citep{borucki2010kepler}, after two of its reaction wheels failed. \citet{giles2018k2} used radial velocity (RV) observations from CORALIE \citep{queloz2000coralie} and the High Accuracy Radial velocity Planet Searcher \citep[HARPS;][]{pepe2000harps} to confirm the planetary nature of K2-140 b and to constrain its physical and orbital properties. 

The planet was independently characterized in \citet{korth2019k2}, which incorporated RV data from the FIbre-fed \'{E}chelle Spectrograph (FIES) \citep{telting2014fies} within their analysis. While \citet{giles2018k2} found a low but nonzero eccentricity for the planet ($e=0.120^{+0.056}_{-0.046}$), \citet{korth2019k2} found that the planet was consistent with $e=0.$ Data from both of these past studies is incorporated within our joint analysis. 

\section{Observations}
\label{section:observations}

We obtained 18 radial velocity measurements of K2-140 with the Keck/HIRES instrument from 9:25-15:40 UT on Feb 24, spanning a full transit of K2-140 b. Conditions were favorable throughout most of the observing period, with typical seeing ranging from $1.0\arcsec-1.3\arcsec$. A spike in humidity during the post-transit baseline observations led to a telescope dome closure at 14:40 UT, resulting in a $\sim$40-minute gap in data before the last radial velocity measurement.

All RV observations were obtained using the C2 decker ($14\arcsec\times0.861\arcsec, R=60,000$) and an iodine absorption cell, which imprints a dense forest of molecular iodine features onto each spectrum to enable high Doppler precision \citep{butler1996attaining}. The $14\arcsec$ length of the C2 decker allows for direct sky subtraction, improving RV precision for faint stars such as K2-140. The median exposure time was $1,119$ seconds, with $\sim$34k exposure meter counts per spectrum. 

Our dataset was reduced using the California Planet Search pipeline outlined in \citet{howard2010california}, and we obtained a typical signal-to-noise ratio of 71 per pixel from the reduced spectra. The HIRES radial velocity results and uncertainties can be found in Table \ref{tab:rv_data} and are shown in the rightmost panel of Figure \ref{fig:rv_joint_fit}. We include the S-index and associated uncertainty at each observation in Table \ref{tab:rv_data} for reference.

We also obtained a 45-minute iodine-free HIRES exposure of K2-140 using the B3 decker ($14.0\arcsec\times0.574\arcsec, R=72,000$) two nights after the measurement of the Rossiter-McLaughlin effect, during UT Feb 26. This template observation was used to calibrate our RVs and to precisely determine stellar parameters (see Section \ref{section:stellar_parameters}). Conditions were favorable during this measurement, and seeing was $1.2\arcsec$. Our reduced template had a signal-to-noise ratio of $123$ per pixel ($106$k exposure meter counts).

\begin{deluxetable}{ccccc}
\tablecaption{HIRES radial velocities for the K2-140 system.\label{tab:rv_data}}
\tabletypesize{\scriptsize}
\tablehead{
\colhead{Time (BJD)} & \colhead{RV (m/s)} & \colhead{$\sigma_{\rm RV}$ (m/s)} & \colhead{S-index} & \colhead{$\sigma_S$}}
\tablewidth{300pt}
\startdata
2459269.903807 & 17.62 & 2.48 & 0.165 & 0.001 \\
2459269.917697 & 1.43 & 2.59 & 0.166 & 0.001 \\
2459269.931633 & 17.18 & 2.55 & 0.160 & 0.001 \\
2459269.944029	& 23.36 & 2.58 & 0.168 & 0.001 \\
2459269.956264	& 23.63 & 2.38 & 0.161 & 0.001 \\
2459269.968059	& 22.22 & 2.47 & 0.166 & 0.001 \\
2459269.978707	& 14.54 & 2.35 & 0.159 & 0.001 \\
2459269.989634	& 11.66 & 2.34 & 0.160 & 0.001 \\
2459270.000977	& -2.85 & 2.31 & 0.165 & 0.001 \\
2459270.013142	& -15.76 & 2.19 & 0.167 & 0.001 \\
2459270.026233	& -14.19 & 2.41 & 0.163 & 0.001 \\
2459270.040065	& -18.89 & 2.45 & 0.160 & 0.001 \\
2459270.054024	& -23.71 & 2.41 & 0.159 & 0.001 \\
2459270.069384	& -10.01 & 2.63 & 0.161 & 0.001 \\
2459270.083054	& 2.645 & 2.96 & 0.155 & 0.001 \\
2459270.097753	& -20.13 & 2.85 & 0.152 & 0.001 \\
2459270.110694	& -14.47 & 3.08 & 0.147 & 0.001 \\
2459270.153058	& -14.61 & 2.71 & 0.155 & 0.001 \\
\enddata
\end{deluxetable}

\section{Obliquity Modeling} 
\label{section:spinorbitmodel}

\begin{figure*}
    \centering
    \includegraphics[width=1.0\linewidth]{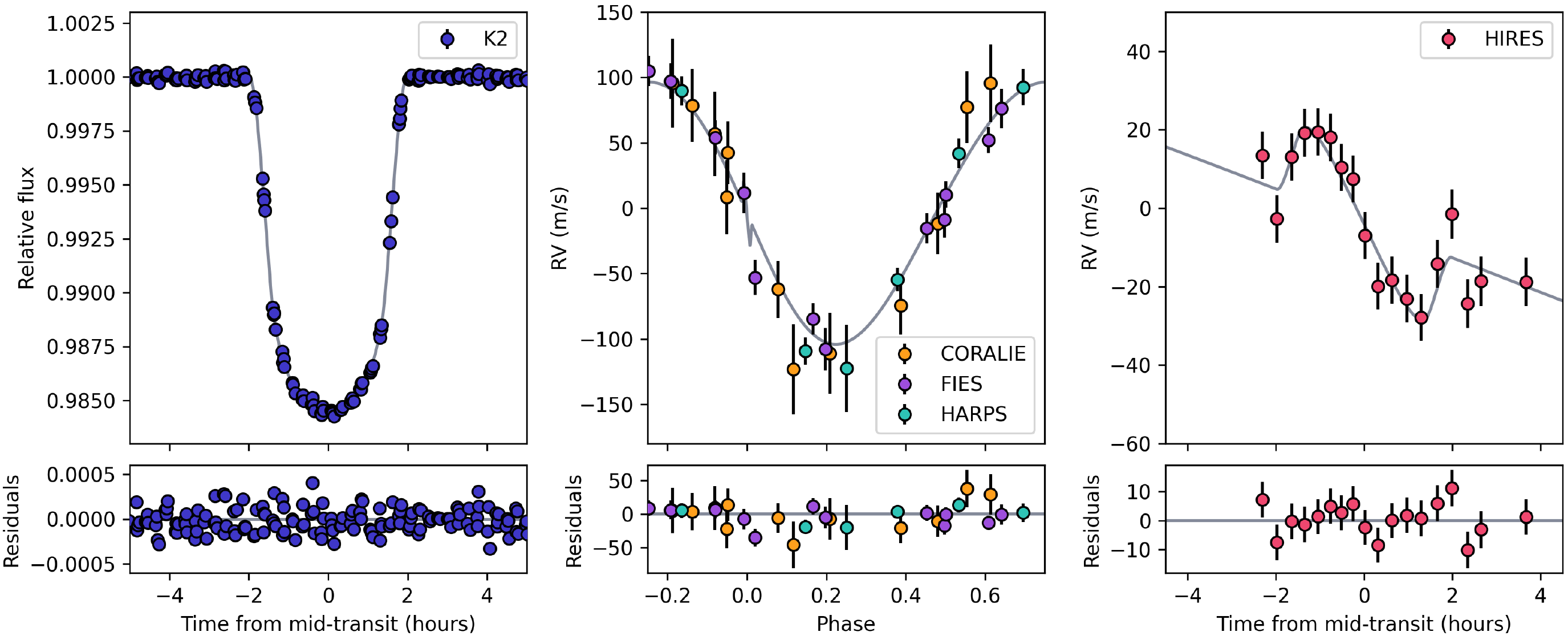}
    \caption{Joint fit to photometry, out-of-transit RV data, and the in-transit Rossiter-McLaughlin RV data obtained for K2-140 b. The model is shown in gray, while data is provided in color with modeled constant offsets and jitter terms included. The associated residuals are provided below each panel.}
    \label{fig:rv_joint_fit}
\end{figure*}

To determine the sky-projected spin-orbit angle $\lambda$ for K2-140\,b, we used the \texttt{allesfitter} Python package \citep{gunther2020allesfitter}  to jointly model the in-transit HIRES radial velocity data together with photometry from $K2$ and archival radial velocity datasets available from the FIES, CORALIE, and HARPS spectrographs.

All fitted parameters listed in Table \ref{tab:system_properties} were allowed to vary, and each parameter was initialized with uniform priors. Initial guesses for $P$, $T_{0}$, $\cos{i}$, $R_{p}/R_{\star}$, $(R_{\star}+R_{p})/a$, $K$, $\sqrt{e}\,\cos{\,\omega}$, and $\sqrt{e}\,\sin{\,\omega}$ were obtained using values from \citet{giles2018k2}. The two limb darkening coefficients $q_1$ and $q_2$ were each initialized with values of 0.5. We accounted for potential radial velocity offsets between each separate spectrograph, with priors bounded by $\pm1000$ m/s. Jitter terms were modeled separately for each instrument and added in quadrature to the instrumental uncertainties. $\lambda$ was allowed to vary between $-180\degr$ and $+180\degr$.

We ran an affine-invariant Markov Chain Monte Carlo (MCMC) analysis with 100 walkers to sample the posterior distributions of all model parameters. The best-fit model parameters and their associated 1$\sigma$ uncertainties were extracted after obtaining 500,000 accepted steps per walker. Our results are listed in Table \ref{tab:system_properties} and are in good agreement with the associated values obtained by \citet{giles2018k2} and \citet{korth2019k2}.

The best-fit joint model is shown in Figure \ref{fig:rv_joint_fit} together with each dataset included in the analysis, as well as the residuals of each fit. The fitted and derived parameters corresponding to this model are provided in Table \ref{tab:system_properties}. We obtain a low but nonzero eccentricity $e=0.069^{+0.042}_{-0.028}$ for K2-140 b, in agreement with the value derived by \citet{giles2018k2}. K2-140 is consistent with alignment, with $\lambda=0.5\degr\pm9.7\degr$ and $v\sin i_* = 2.51\pm0.38$ km/s.

\begin{deluxetable*}{llllll}
\tablecaption{System properties derived for K2-140.\label{tab:system_properties}}
\tabletypesize{\scriptsize}
\tablehead{\colhead{Parameter} & \colhead{Description} & \colhead{Priors} & \colhead{Value} &  \colhead{$+1\sigma$} & \colhead{-1$\sigma$}}
\tablewidth{300pt}
\startdata
\vspace{-2mm}
& & & & & \\
Fitted Parameters: & & & & & \\
$R_p / R_\star$\dotfill & Planet-to-star radius ratio\dotfill & $\mathcal{U}(0;1)$\tablenotemark{$*$} & $0.1163$ & 0.0012 & 0.0011 \\
$(R_\star + R_p) / a$\dotfill & Sum of radii divided by the orbital semimajor axis\dotfill & $\mathcal{U}(0;1)$ & $0.0960$ & 0.0043 & 0.0039 \\
$\cos{i}$\dotfill & Cosine of the orbital inclination\dotfill & $\mathcal{U}(0;1)$ & $0.0518$ & 0.0053 & 0.0048 \\
$T_{0}$\dotfill & Mid-transit epoch ($\mathrm{BJD}$)\dotfill & $2458435.7138$\tablenotemark{$\dagger$}          &  $2458429.1426$ & 0.0015 & 0.0015 \\
$P$\dotfill & Orbital period (days)\dotfill &$\mathcal{U}(5.569188;7.569188)$ & $6.569199$ & 1.2e-05 & 1.2e-05 \\
$K$\dotfill & Radial velocity semi-amplitude ($\mathrm{m/s}$)\dotfill & $\mathcal{U}(0;1000)$  & $106.5$ & 4.7 & 4.7 \\
$\sqrt{e} \cos{\omega}$\dotfill & Eccentricity parameter 1\dotfill & $\mathcal{U}(-1.0;1.0)$  & $-0.156$ & $0.067$ & $0.053$ \\
$\sqrt{e} \sin{\omega}$\dotfill & Eccentricity parameter 2\dotfill & $\mathcal{U}(-1.0;1.0)$  & $0.189$ & $0.098$ & $0.15$ \\
$q_{1}$\dotfill & Quadratic limb darkening coefficient 1\dotfill & $\mathcal{U}(0.0;1.0)$ &  $0.74$ & $0.15$ & $0.14$ \\
$q_{2}$\dotfill & Quadratic limb darkening coefficient 2\dotfill & $\mathcal{U}(0.0;1.0)$ &  $0.175$ & $0.072$ & $0.055$ \\
$\Delta_{\rm RV, FIES}$\dotfill & RV offset, FIES (m/s)\dotfill & $\mathcal{U}(-1000.0;1000.0)$ & $1.1287$ & 0.0044 & 0.0042 \\
$\Delta_{\rm RV, CORALIE}$\dotfill & RV offset, CORALIE (m/s)\dotfill & $\mathcal{U}(-1000.0;1000.0)$ & $1.2141$ & 0.0083 & 0.0083 \\
$\Delta_{\rm RV, HARPS}$\dotfill & RV offset, HARPS (m/s)\dotfill & $\mathcal{U}(-1000.0;1000.0)$ & $1.2459$ & 0.0053 & 0.0053 \\
$\Delta_{\rm RV, HIRES}$\dotfill & RV offset, HIRES (m/s)\dotfill & $\mathcal{U}(-1000.0;1000.0)$ & $0.0051$ & 0.0034 & 0.0034 \\
$\lambda$\dotfill & Sky-projected spin-orbit angle ($\degr$)\dotfill &  $\mathcal{U}(-180.0;180.0)$ & $0.5$ & $9.7$ & $9.7$ \\
\vspace{2mm}
$v\sin i_*$\dotfill & Sky-projected stellar rotational velocity (km/s)\dotfill & $\mathcal{U}(0.0;20.0)$ & $2.51$ & $0.38$ & $0.38$ \\
Derived Parameters: & & & & & \\
$R_\mathrm{p}$  & Planetary radius ($\mathrm{R_{J}}$)            &...         & $1.203$       & $0.076$    & $0.076$        \\
$M_\mathrm{p}$  & Planetary mass  ($\mathrm{M_{J}}$)             &...         & $1.13$        & $0.12$     & $0.11$         \\
$b$             & Impact parameter                                 &...         & $0.574$       & $0.022$    & $0.022$        \\
$T_\mathrm{14}$ & Transit duration     (h)                         &...         & $3.946$       & $ 0.016$   & $ 0.016$       \\
$\delta$        & Transit depth                                    &...         & $0.015490$    & $4.2e-05$ & $4.2e-05$     \\
$a$ & Semimajor axis (au) & ... & 0.0575 & 0.0046 & 0.0043 \\
$i$             & Inclination  ($\degr$)                        &...         & $87.03$       & $0.27$     & $0.30$          \\
$e$             & Eccentricity                                     &...         & $0.069$       & $0.042$    & $0.028$        \\
$\omega$        & Argument of periastron  ($\degr$)                    &...         & $131$         & $38$       & $20$           \\
$u_\mathrm{1}$  & Limb darkening parameter 1                                   &...         & $0.300$       & $0.088$    & $0.080$        \\
$u_\mathrm{2}$  & Limb darkening parameter 2                                  &...         & $0.56$        & $0.15$     & $0.16$ 
\enddata
\tablenotetext{$*$}{$\mathcal{U}(a;b)$ is a uniform prior with lower and upper limits $a$ and $b$, respectively.}
\tablenotetext{\dagger}{We provided a reference value for $T_{0}$. During the fit, \texttt{allesfitter} can shift epochs to the data center to derive an optimal $T_{0}$.}
\end{deluxetable*}

\section{Stellar Parameters} 
\label{section:stellar_parameters} 

An understanding of host star properties can help to contextualize the evolutionary pathways through which a system may have reached its current state. We extracted stellar parameters from our Keck/HIRES template spectrum of K2-140 using the data-driven spectroscopic modeling program \textit{The Cannon} \citep{ness2015cannon, casey2016cannon}, following the methods of \citet{rice2020stellar}. 

Given a set of uniformly processed input training spectra and associated stellar ``labels" -- that is, stellar parameters and elemental abundances -- \textit{The Cannon} constructs a generative model describing the probability density function of flux at each wavelength as a function of the labels. The model can then be applied to a new set of spectra, uniformly processed in the same manner as the training set, to obtain the associated stellar labels. 

We trained \textit{The Cannon} using the uniformly analyzed Spectral Properties of Cool Stars (SPOCS) catalogue \citep{brewer2016spectral} of 18 stellar labels, including 3 global stellar parameters ($T_{\rm eff}$, $\log g$, $v\sin i_*$), and 15 elemental abundances: C, N, O, Na, Mg, Al, Si, Ca, Ti, V, Cr, Mn, Fe, Ni, and Y. The sample of 1202 Keck/HIRES spectra vetted in \citet{rice2020stellar} was applied as our training/test set.

\begin{figure}
    \centering
    \includegraphics[width=1.0\linewidth]{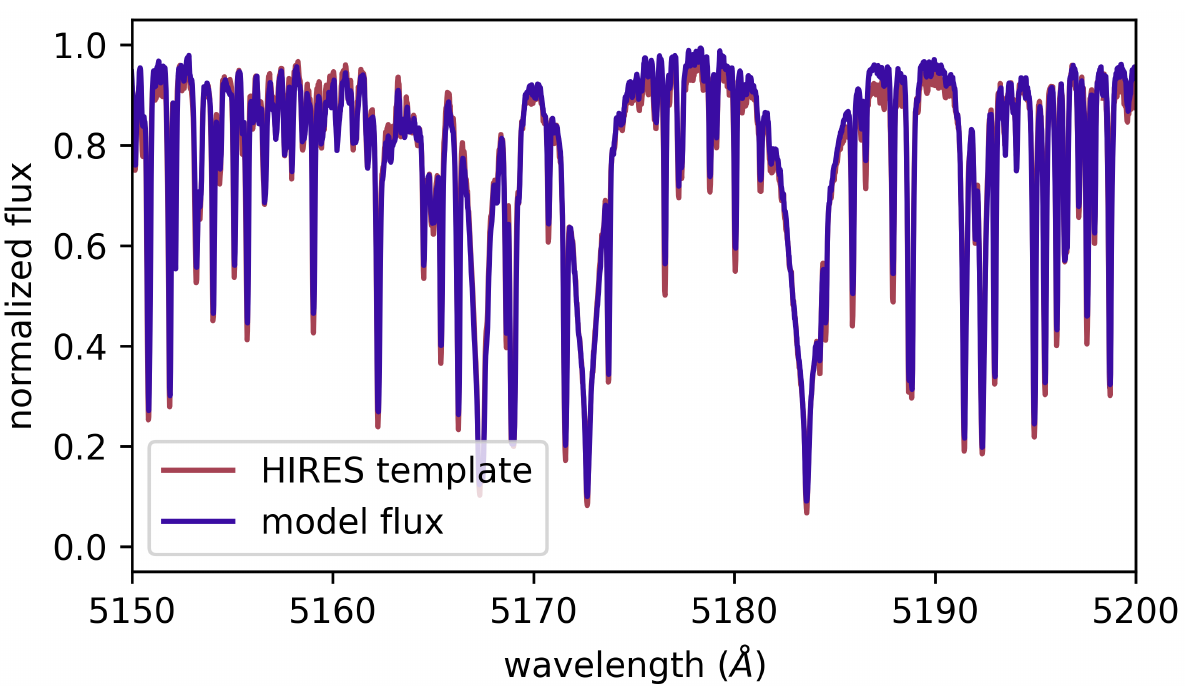}
    \caption{Sample segment of the model spectrum returned by \textit{The Cannon}, shown alongside the K2-140 HIRES template spectrum in the vicinity of the Mg Ib triplet (lines at 5167, 5172, and 5183 \AA).}
    \label{fig:spectroscopy_model_comparison}
\end{figure}

The continuum baseline of each spectrum was uniformly fit and divided out using the iterative polynomial fitting procedure outlined in \citet{valenti2005spectroscopic}. Then, we split the SPOCS sample into an 80\%/20\% training/test split, applied the telluric mask from \citet{rice2020stellar} to all spectra, and trained the model, using the scatter of the test set results to determine the uncertainties of each extracted parameter. Finally, the trained model was applied to the newly acquired K2-140 Keck/HIRES template spectrum. A segment of the obtained model spectrum is shown in comparison with the HIRES template data in Figure \ref{fig:spectroscopy_model_comparison}.

Our results are provided in Table \ref{tab:stellar_properties} together with estimates from previous works for reference. Our values for $T_{\rm eff}$, $\log g$, and [Fe/H] are in agreement with those acquired in previous studies. The $v\sin i_*$ value obtained using \textit{The Cannon} is lower than previous estimates but is in agreement with the RM joint fit results and \citet{korth2019k2} within $1\sigma$. We find that K2-140 is metal-enriched relative to solar abundances, consistent with past evidence showing that short-period giant planets are more common around metal-rich stars \citep{fischer2005planet}.

\begin{deluxetable*}{llllll}
\tablecaption{Stellar parameters for K2-140.\label{tab:stellar_properties}}
\tabletypesize{\scriptsize}
\tablehead{
\multicolumn{1}{p{1.6cm}}{\raggedright Parameter \\} & \multicolumn{1}{p{1.6cm}}{\raggedright Unit \\} & \multicolumn{1}{p{1.6cm}}{\raggedright The Cannon \\ (this work) \\} & \multicolumn{1}{p{1.6cm}}{\raggedright RM joint fit \\ (this work) \\} & \multicolumn{1}{p{1.6cm}}{\raggedright Giles+ 2019} & \multicolumn{1}{p{1.6cm}}{\raggedright Korth+ 2019 \\}}
\tablewidth{300pt}
\startdata
$T_{\rm eff}$ & K & $5610\pm59$ & - & $5654\pm55$ & $5585\pm120$ \\
log$g$ & cm/s$^2$ & $4.4\pm0.1$ & - & 4.452$^{+0.010}_{-0.009}$ & $4.4\pm0.2$ \\
$v\sin i_*$ & km/s & $2.31\pm1.07$ & $2.51\pm0.38$ & $3.8\pm0.2$ & $3.6\pm1.0$ \\
$[\rm{Fe/H}]$ & dex & 0.18$\pm0.04$ & - & $0.12\pm0.045$ & $0.10\pm0.10$ \\
$[\rm{C/H}]$ & dex & $0.14\pm0.08$ & - & - & - \\
$[\rm{N/H}]$ & dex & $0.05\pm0.09$ & - & - & - \\
$[\rm{O/H}]$ & dex & $0.06\pm0.09$ & - & - & - \\
$[\rm{Na/H}]$ & dex & $0.13\pm0.07$ & - & - & $0.12\pm0.10$ \\
$[\rm{Mg/H}]$ & dex & $0.12\pm0.04$ & - & - & $0.27\pm0.10$ \\
$[\rm{Al/H}]$ & dex & $0.22\pm0.13$ & - & - & - \\
$[\rm{Si/H}]$ & dex & $0.18\pm0.05$ & - & - & - \\
$[\rm{Ca/H}]$ & dex & $0.24\pm0.04$ & - & - & $0.12\pm0.10$ \\
$[\rm{Ti/H}]$ & dex & $0.24\pm0.05$ & - & - & - \\
$[\rm{V/H}]$ & dex & $0.17\pm0.06$ & - & - & - \\
$[\rm{Cr/H}]$ & dex & $0.19\pm0.05$ & - & - & - \\
$[\rm{Mn/H}]$ & dex & $0.18\pm0.06$ & - & - & - \\
$[\rm{Ni/H}]$ & dex & $0.18\pm0.05$ & - & - & $0.20\pm0.10$ \\
$[\rm{Y/H}]$ & dex & $0.23\pm0.12$ & - & - & - \\
\enddata
\end{deluxetable*}

\begin{figure*}
    \centering
    \includegraphics[width=1.0\linewidth]{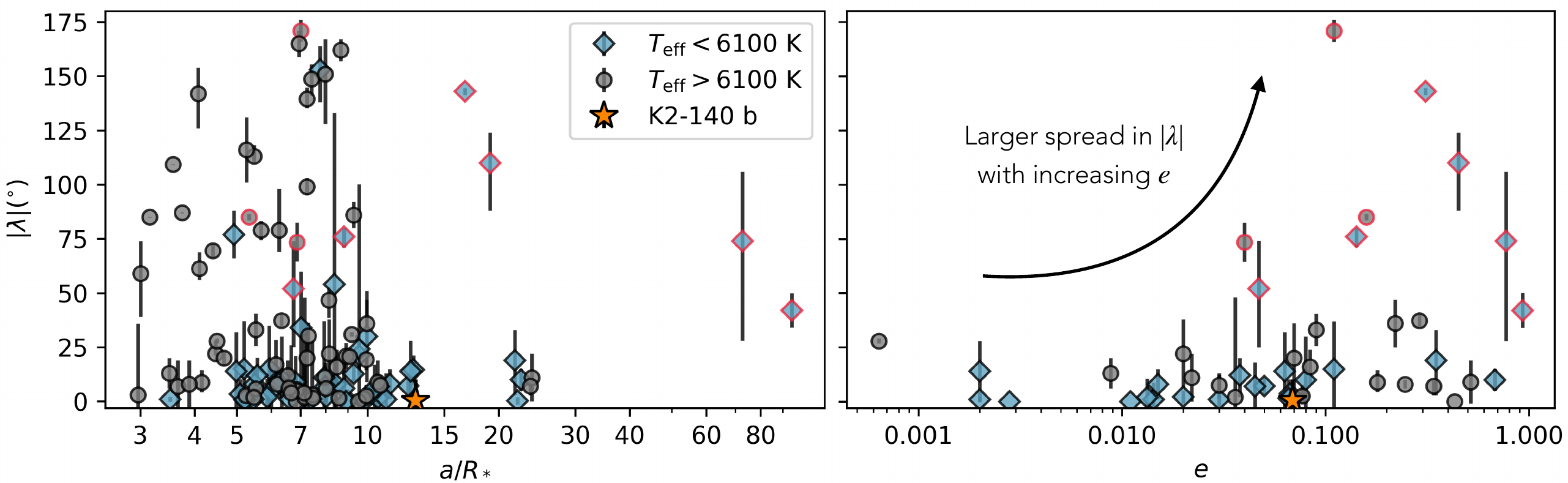}
    \caption{\textit{Left:} K2-140 b relative to other Rossiter-McLaughlin measurements of giant planets ($M>0.3M_J$) in the TEPCat catalogue \citep{southworth2011homogeneous}, with $a$ and $R_*$ obtained through cross-matching with the NASA Exoplanet Archive and missing values filled in from the Extrasolar Planets Encyclopaedia. In $a/R_*$ space, K2-140 b lies just exterior to most planets with measured spin-orbit angles. This makes it one of only a handful of planets with a tidal realignment timescale $\tau >>$ stellar age, such that it would not have had time to realign if it had been misaligned at the time of protoplanetary disk dispersal. \textit{Right:} Distribution of measured obliquities as a function of eccentricity for planets with $e>0$. The range of measured $\lambda$ values is larger at high $e$ for planets orbiting stars both above and below the Kraft break. Planets with $|\lambda|>40\degr$ and $e>0.01$ are outlined in red in both panels. Among the planets with $e\neq0$, the spread in misalignments is larger at higher eccentricities.}
    \label{fig:a_over_Rstar_v_lambda_withecc_K2140}
\end{figure*}

\section{Discussion}
\label{section:discussion} 

\subsection{Implications of the Low \texorpdfstring{$\lambda$}{} of K2-140 b}

At $T_{\rm eff}=5585$ K, K2-140 should have a convective envelope. In the framework of equilibrium tides, the planet's timescale for realignment from turbulent friction would thus follow 
\begin{equation}
    \tau_{CE} = \frac{10^{10} \rm{yr}}{(M_p/M_*)^2}\Big(\frac{a/R_*}{40}\Big)^{6},
\label{eq:tau}
\end{equation}
where $\tau_{CE}$ is the realignment timescale for host stars with convective envelopes, and $M_p/M_*$ is the planet-to-star mass ratio \citep{zahn1977tidal, albrecht2012obliquities}. For K2-140\,b, $\tau_{CE}=1.2\times10^{13}$ yr, longer than the age of the Universe. As a result, we conclude that K2-140\,b was likely aligned at the time of protoplanetary disk dispersal.

The theory of equilibrium tides presented in Equation \ref{eq:tau} is employed as a simplified heuristic for a broader theoretical framework that has substantially advanced over recent years \citep{ogilvie2014tidal}. A key problem in the equilibrium tides framework is that, under standard assumptions, hot Jupiters should experience rapid orbital decay. \citet{lai2012tidal} demonstrated that one component of the tidal potential (the ``obliquity tide") can excite inertial waves in the convective envelopes of cool stars, which, when damped, can enhance dissipation of the stellar obliquity without shrinking the companion's orbit. Obliquity tides have been further explored in additional work \citep{ogilvie2013tides, lin2017tidal, anderson2021possible}. While obliquity tides are not included in our analysis, previous work has demonstrated that Equation \ref{eq:tau} is a useful heuristic revealing that low-obliquity systems tend to have shorter tidal timescales than high-obliquity systems \citep{albrecht2012obliquities}.

K2-140\,b is one of only 13 giant planets with a measured spin-orbit angle at $a/R_*>12$, as shown in the left panel of Figure \ref{fig:a_over_Rstar_v_lambda_withecc_K2140}. Of these planets, it is one of only a few with an aligned orbit despite its long tidal realignment timescale. The alignment of K2-140\,b suggests that at least some hot Jupiters form through quiescent pathways, such as in-situ formation or disk migration in an initially aligned disk. 

The highly misaligned planets at large $a/R_*$ each have high eccentricities (right panel of Figure \ref{fig:a_over_Rstar_v_lambda_withecc_K2140}), indicative of strong dynamical interactions that may have produced both elevated obliquities and eccentricities. The coexistence of this population with the dynamically quiescent K2-140 system at large $a/R_*$ suggests that there are multiple hot Jupiter formation channels.

\begin{figure}
    \centering
    \includegraphics[width=1.0\linewidth]{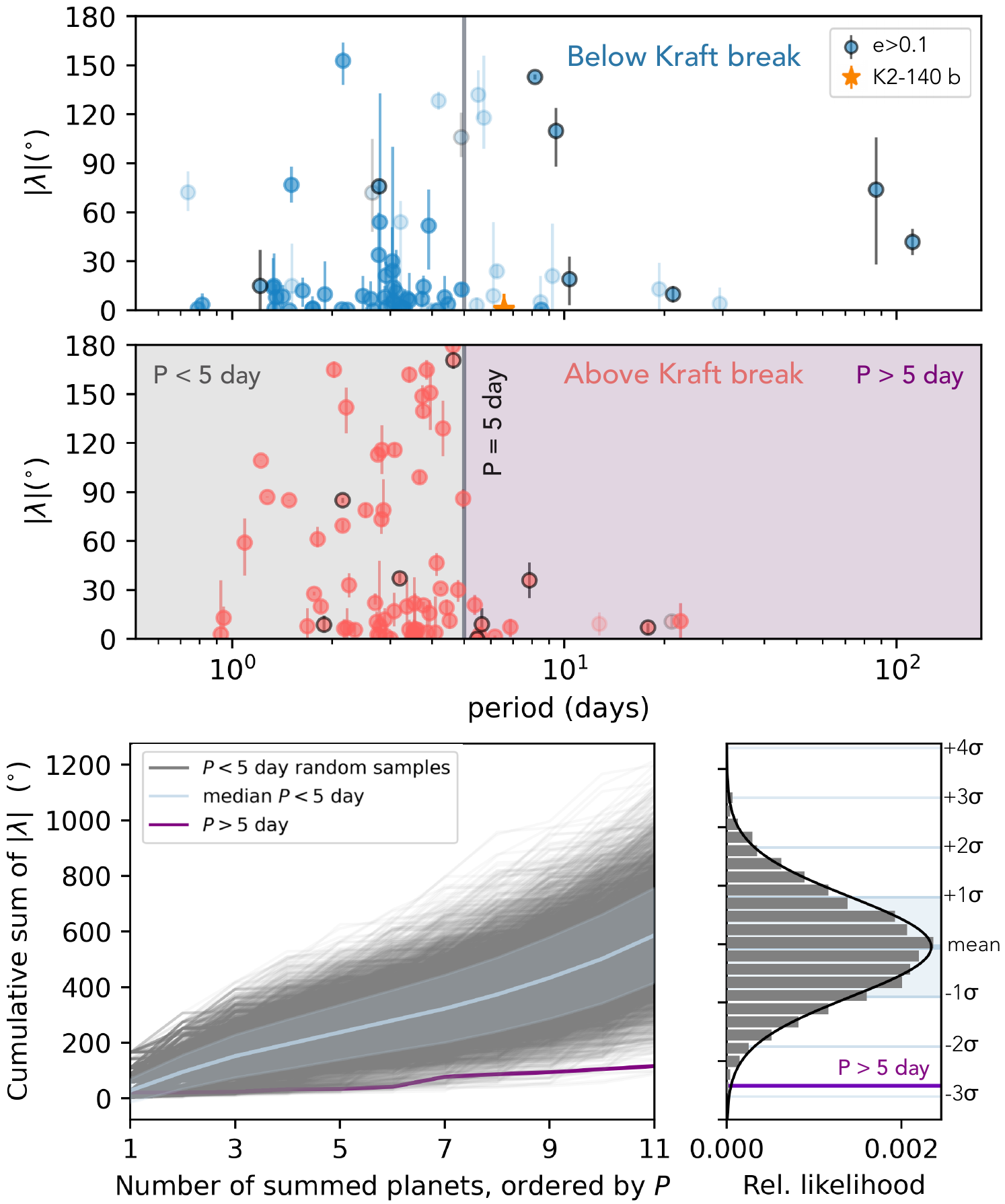}
    \caption{Obliquity distribution of all stars below (top panel) and above (middle panel) the Kraft break with measured $\lambda$ values in the TEPCat catalogue, provided as a function of the companion planet's orbital period. Measurements for companion planets with $e>0.1$ are bordered in black. Measurements for companion planets with $M<0.3M_J$ are shown at lower opacity. The cumulative sums in the bottom panel include only planets above the Kraft break, comparing the $P>5$ day population (purple) with 5000 $P<5$ day samples (gray), randomly sampled without replacement. The histogram on the bottom right shows a vertical cut through the final cumulative sum distribution. The $P>5$ day cumulative sum is a 2.79$\sigma$ outlier from the random draw distribution.}
    \label{fig:period_distribution}
\end{figure}

No neighboring planets have yet been found in the K2-140 system, despite previous observations showing that inner, coplanar companions are common for slightly longer-period warm Jupiters \citep[$10<P<200$ days;][]{huang2016warm} and predictions that outer, mutually-inclined companions with $P\lesssim 100$ days should be regularly produced by in-situ hot Jupiter formation via core accretion \citep{batygin2016situ}. However, additional low-mass or distant planets in the system cannot be ruled out by the existing observations, with radial velocity residuals of tens of m/s and an observing baseline of only $\sim$4 years.

\subsection{Motivation for Additional Obliquity Measurements in Tidally Detached Systems through SOLES}

Existing Rossiter-McLaughlin measurements of wide-orbiting planets suggest that this population may have an intrinsically different spin-orbit angle distribution from the shorter-period hot Jupiter population. In particular, the trend of systematically lower obliquities observed for stars at temperatures below the Kraft break \citep[ $T_{\mathrm{eff}}\approx6100$ K;][]{kraft1967studies, winn2010hot, schlaufman2010evidence}, which may result from tidal damping \citep[e.g.][]{wang2021aligned}, is not immediately evident at longer orbital periods, as shown in Figure \ref{fig:period_distribution}. 

Conversely, Figure \ref{fig:period_distribution} reveals tentative ($2.79\sigma$) evidence that relatively long-period planets ($P>5$ days) around hot stars are preferentially \textit{less} misaligned than their shorter-period counterparts. Furthermore, the most misaligned $P>5$ day planet in the middle panel, KELT-6\,b \citep[$P=7.8$ days, $|\lambda|=36\pm11\degr$;][]{damasso2015gaps}, orbits a star directly bordering the nominal Kraft break \citep[$T_{\rm eff}=6102$\,K;][]{collins2014kelt}, making it an ambiguous member of its group. We note that hot stars typically have larger radii and thus stronger tidal dissipation at a given orbital period; however, a similar trend has also been previously suggested for a smaller population of planets at large $a/R_*$ \citep{yu2018epic}. 

Longer-period planets have comparatively long tidal alignment timescales, especially around hot stars. The small spin-orbit angles observed for exoplanets around hot stars, therefore, may suggest that protoplanetary disks tend to be aligned at the time of gas dispersal \citep[in contrast with a primordial spin-orbit misalignment from a tilted disk;][]{batygin2012primordial, spalding2015magnetic}. In this case, hot Jupiters would need to obtain their misalignments after the protoplanetary disk has dispersed, favoring high-eccentricity migration as an important formation mechanism \citep[e.g.][]{wu2003planet, fabrycky2007shrinking}. Most of the misaligned planets at $P>5$ days are on eccentric orbits, suggesting that interactions with stellar or planetary companions in the systems may have induced both elevated obliquities and eccentricities. 

If a growing sample of spin-orbit angles at $P>5$ days reveals that cool stars below the Kraft break are preferentially more misaligned than hotter stars above the Kraft break, this may also be suggestive of interactions with additional companions. In the presence of an external, inclined Jovian-mass companion, spin-orbit misalignments can be excited preferentially in cool star systems due to a secular resonance between the host star's spin axis precession frequency and nodal precession induced by interactions with the companion \citep{anderson2018teetering}. Cool stars spin down over time due to magnetic braking, enabling this resonant excitation. Hot stars, by contrast, lack a convective envelope and continue to rapidly rotate over their lifetimes. As a result, irrespective of the external giant planet companion rate around hot stars, this mechanism should occur only in cool star systems. 

The substantially inclined companions necessary to induce this mechanism would also be capable of exciting the inner planet's orbital eccentricity. This may be reflected by the apparent increase in both misalignments and eccentricities for long-period planets around cool stars, shown in the top panel of Figure \ref{fig:period_distribution}. Conversely, the alignment of K2-140 b and similar planets on low-eccentricity orbits indicates that they should not have nearby giant planet companions with large ($>10\degr$) mutual inclinations within $\sim2$ au.

If the trend of large misalignments for cool stars hosting tidally detached planets persists, while misalignments remain small for their hot star counterparts, further monitoring would be warranted to constrain the long-period giant planet companion rate for misaligned cool star systems. Additional observations are needed to parse the emerging relationship between stellar temperature, eccentricity, and obliquity for wide-separation planets.



\section{Acknowledgements}
\label{section:acknowledgements}

We thank Konstantin Batygin and Kassandra Anderson for helpful discussions that have refined this work. We also thank the anonymous referee for their helpful suggestions that have improved the quality of this manuscript. M.R. is supported by the National Science Foundation Graduate Research Fellowship Program under Grant Number DGE-1752134. The data presented herein were obtained at the W. M. Keck Observatory, which is operated as a scientific partnership among the California Institute of Technology, the University of California and the National Aeronautics and Space Administration. The Observatory was made possible by the generous financial support of the W. M. Keck Foundation. This work is supported by Astronomical Big Data Joint Research Center, co-founded by National Astronomical Observatories, Chinese Academy of Sciences and Alibaba Cloud. This research has made use of the Keck Observatory Archive (KOA), which is operated by the W. M. Keck Observatory and the NASA Exoplanet Science Institute (NExScI), under contract with the National Aeronautics and Space Administration. This research has made use of the NASA Exoplanet Archive, which is operated by the California Institute of Technology, under contract with the National Aeronautics and Space Administration under the Exoplanet Exploration Program.

\software{\texttt{numpy} \citep{oliphant2006guide, walt2011numpy, harris2020array}, \texttt{matplotlib} \citep{hunter2007matplotlib}, \texttt{pandas} \citep{mckinney2010data}, \texttt{scipy} \citep{virtanen2020scipy}, \texttt{allesfitter} \citep{gunther2020allesfitter}, \texttt{emcee} \citep{foremanmackey2013}}

\facility{Keck: I (HIRES), Exoplanet Archive, Extrasolar Planets Encyclopaedia}

\bibliography{bibliography}
\bibliographystyle{aasjournal}

\end{document}